\documentclass{article}

\usepackage[utf8]{inputenc}
\usepackage{graphicx}
\usepackage{caption}
\usepackage{verbatim}
\usepackage{float}
\usepackage{multirow}
\usepackage{tabularx}
\usepackage{longtable}
\usepackage{etex}
\usepackage{enumitem}
\usepackage{lscape}
\usepackage{booktabs}

\usepackage{colortbl}

\title{Automated UX Insights from User Research Videos by Integrating Facial Emotion and Text Sentiment}

\author{Simran Kaur Ghatoray  and Yongmin Li\\
Department of Computer Science, Brunel University London}
\date{}

\begin{document}
\maketitle

\abstract{
Emotion recognition technology has been studied from the past decade. With its growing importance and applications such as customer service, medical, education, etc., this research study aims to explore its potential and importance in the field of User experience evaluation. Recognizing and keeping track of user emotions in user research video is important to understand user needs and expectations from a service/product. Little research has been done that focuses on automating emotion extraction from a video where more than one modality has been incorporated in the field of UX. 
The study aims at implementing different modalities such as facial emotion recognition, speech-to-text and text-based emotion recognition for capturing emotional nuances from a user research video and extract meaningful actionable insights.
For selection of facial emotion recognition model, 10 pre-trained models were evaluated on three benchmark datasets i.e. FER-2013, AffectNet and CK+, selecting the model with most generalization ability. To extract speech and convert to text, OpenAI’s Whisper model was implemented and finally the emotions from text were recognized using a pre-trained model available at HuggingFace website having an evaluation accuracy more than 95\%. The study also integrates the gathered data using temporal alignment and fusion for deeper and contextual insights. The study further demonstrates a way of automating data analysis through PandasAI Python library where OpenAI’s GPT- 4o model was implemented along with a discussion on other possible solutions. This study is an attempt to demonstrate a proof of concept where automated meaningful insights are extracted from a video based on user emotions.}


\section{Introduction}
In today’s digital world, user emotions play a vital role in developing user-centric designs. The emotional state of a person can affect concentration, decision-making skills, and task-solving skills \cite{Kołakowska2014}. The growing importance of recognizing emotions can be seen in different fields such as healthcare, medical, customer service, human-computer interactions, education, gaming, etc. Among its vast applications, in recent years, focus has been made towards User experience and its evaluation methods. \par

User experience (UX) is a growing field that includes disciplines such as User Interface Design. Better user design leads to greater user satisfaction following good revenue and retention \cite{Souza2022}. UX evaluation is carried out to ensure better user satisfaction and meet user requirements through various methods and techniques. Usability and UX are intertwined terms, but serve different purposes. UX methods focus on improving user satisfaction by achieving pragmatic and hedonic goals, while usability methods aim to improve human performance \cite{Souza2022}. With the advancement in the field of Artificial Intelligence, there always remains a question of how AI methods and techniques can improve UX evaluation? \par

Different techniques, methods and modalities, now, make it possible to capture the whole picture of user emotions, allowing to change the traditional ways of UX evaluation. User emotions being a crucial part for the betterment of systems, can be detected and recognized with the help of machine learning and deep learning methods. Various modalities such as facial emotion recognition, speech emotion recognition, text based emotion recognition allow one to automate the process of analyzing user emotions be it in an image or video, revolutionizing the ways of user feedback analysis. \par

Researchers have made efforts to implement AI-driven solutions to make UX evaluation easy and efficient \cite{Souza2022}\cite{Aviz2019}. The literature on improving individual elements such as facial recognition in UX by \cite{Afriansyah2021}, emotion detection using EEG signals \cite{Gannouni2023}, eye tracking and tracking of mouse clicks, etc. represented by \cite{Souza2022}\cite{Qu2017}\cite{Novák2023} could be found. Some work in building a framework for UX evaluation by \cite{Drungilas2024}\cite{Cordeiro2024} can be seen. \par

Emotions are complex, and it might not be possible to grasp the intentions and feelings of the user through a single modality of recognizing emotions. Integration of data gathered through the combination of different modalities such as face, speech, etc. provide deeper and meaningful insights. It also helps to capture any anomalies detected in the video. As demonstrated by \cite{Cordeiro2024}, the integration of facial and speech emotion recognition methods can be a powerful tool to improve UX.\par

Among all the applications of emotion recognition technology, little research and solutions have been provided focusing on user emotions in an end-to-end workflow. The practice of manually analyzing the change and evolution of user emotions while reviewing a service/product in a recorded session still remains. UX designers re-watch the whole session number of times to keep track of emotion nuances, limiting the insights available to them. This study, focusing on capturing user emotion nuances in user research videos through combination of different modalities, demonstrates a concept, one way of solving the problem of collecting meaningful insights from a video, making the end-to-end process of UX evaluation automated. \par

The project aims to extract meaningful and actionable information from user research videos, focusing on the use of facial, speech-to-text, and text-based emotion recognition modalities for data gathering and integration to capture nuances in user emotions expressed in the video. Following the effective emotion recognition and data integration process, the project aims to provide one of the possible ways to automate the data processing step, to extract and generate insights, through the implementation of a conversational analyst tool. \par

\section{Literature Review}
Emotion Recognition (ER) has been studied for more than a decade. Various methods have been introduced to recognize emotions. Some of which include recognizing emotions from face, speech, tone, text, physiological signals, body language, etc. In \cite{Saxena2020}, extensive research is conducted on various techniques for recognizing emotions through artificial intelligence (AI) from the past decade. The paper covers four modalities for ER, namely face, text, audio and physiological signals, where the experiments have shown that majority of the work has been done in facial, followed by textual and audio emotion recognition. Knowing these technologies and their better use could be an efficient way to handle complex problems on healthcare, media, customer service, and especially Human-Computer Interaction (HCI). \par

Among all the applications, its importance in User experience (UX) research and usability testing needs to be emphasized. UX evaluation (User Testing) is a process to gain insight into user satisfaction with using / reviewing a product or service. It analyzes how well the product has met the expectations of the user. Their valuable response helps in identifying strengths, weakness, and areas of improvement. According to \cite{Novák2023}, the user experience combines the physical and technical aspects of the product with the cognitive processes of the user, focusing on the emotional impacts and satisfaction during the interaction with the product, while usability tests usually focus on the performance of tasks such as the execution time of a task, the number of clicks, etc. \cite{Greenberg2009}. Various methods have been discovered to evaluate the UX process. In \cite{Novák2023}, the author states that UX is usually measured quantitatively on objective data at its core or qualitatively, where usability testing is considered at its core. The author of \cite{Greenberg2009} has reported nearly 96 methods for UX evaluation through comprehensive research, emphasizing the difference between usability testing and UX evaluation. According to the author, the relationship between UX and usability testing is intertwined, but objective usability testing is not a sufficient measure for subjective UX evaluation, as it focuses on how the user feels about a system/service. \par

Many applications of AI technology in the field of UX have focused on usability testing where they keep track of eye, keyboard input, number of mouse clicks, etc. Work has been done in objective evaluation of UX with very little focus on subjective part. In \cite{Souza2022}, a framework is presented for UX evaluation focusing on various tracking techniques such as eye and mouse tracking, keyboard inputs, self-assessment questionnaire to categorize users in terms of performance profile. \cite{Qu2017} proposed an eye tracking technology to objectively evaluate UX for smartphone APPs. 

\subsection{Emotion models and theories}
Understanding emotion models and theories helps contextualize implementation. For decades, psychologists and researchers have proposed different theories and models related to human emotion. In \cite{Ong2015}, model of a lay theory of emotions explains how the observer, called an agent by the lay theory, infers about the target of reasoning. The presence of emotional stimuli and the interaction of these stimuli with other mental states such as goals, generate emotions within the agent. External manifestations of these emotions include speech, body language, facial expressions, and future actions. In \cite{Lopatovska2011}, the emotion theories are categorized into two different views as manifestation and structure.\par

The author states that emotional reactions can arise from either cognitive judgment or bodily responses when focusing on the manifestation point of view, whereas the structural point of view follows discrete and continuous approaches. The discrete approach as a consideration of universally recognized basic emotions (such as fear, anger, disgust, happiness, sadness, and surprise) is described in \cite{Ekman1992}. The continuous approach considers two or more dimensions that describe different emotions, as can be seen in the circumplex model of affect proposed by \cite{Russell1980}. The model distributes the emotions in a two-dimensional circular space containing arousal and valence. The vertical/horizontal axes represent the arousal/valence, where the center of the circle represents the medium level of arousal and the neutral level of valence. In addition to this, six basic emotions on the border of Russell’s proposed model are added in \cite{Fernández2016}.

\subsection{Emotion Recognition Modalities}
A facial recognition (FR) system automatically detects and identifies human faces from a digital image or video frames from a video source \cite{li2001dynamic, Sharma2013,li2001video,li2003constructing}. Researchers approach the problem of facial recognition system in different ways. In \cite{Kortli2020}, the researcher presents state of the art of existing facial recognition techniques through three different approaches such as local – uses features with partially defined face such as Local Binary pattern (LBP), holistic – uses features which describes complete face as a model such as Principal Component Analysis (PCA), Eigenfaces, Support Vector Machine (SVM), Convolutional Neural Network (CNN) and hybrid – combination of local and holistic \cite{li2000multi,li2001constructing,li2003recognising}, whereas in \cite{Setiowati2017}, the researcher divides the solution to the FR problem into two categories of non-deep learning methods– Eigenface, Fisherface, SVM, LBP and deep learning (DL) methods – Multi-layer Perceptron (MLP) and CNN. 

Similar to author’s approach in \cite{Setiowati2017}, researchers in \cite{Canal2022}, \cite{Ko2018},  \cite{Moolchandani2021} also categorize facial emotion recognition (FER) methods into the conventional and DL approach. The general steps involved in developing a FER model includes data pre-processing, feature extraction, classification and finally results and validation. Three types of data pre-processing steps, ie, gray scale conversion, face detection, and dimensionality reduction, are presented by \cite{Canal2022}. Whether a conventional approach or DL is being used, it is usually decided by the methods used for feature extraction and classification. Researchers have different opinions about which approach works best. According to \cite{Kortli2020}, local feature techniques are better in terms of complexity, rotation and accuracy, whereas in \cite{Setiowati2017} experimentation concludes that DL methods are more promising for facial recognition reporting an accuracy of 94.67\% of low-high complexity for DL methods and 90.6\% of low complexity for non-DL methods.\par

Moving forward to other emotion recognition technologies for speech and text, researchers have made an impressive contribution. A Speech Emotion Recognition (SER) system extracts and classifies the existing emotions of the target speaker from a pre-processed speech signal. A comprehensive survey on various SER methods is demonstrated in \cite{Wani2021}. SER systems also follow the same workflow of data pre-processing, feature extraction, classification, and evaluation. Different ways of speech processing such as framing, windowing, normalization, noise reduction, etc. Different classifiers such as Artificial Neural Networks (ANN), KNN, SVM, deep neural networks (DNN), recurrent neural networks (RNN), CNN, etc. are broadly discussed in \cite{Wani2021}. Another technology that could be used is speech-to-text (STT), which recognizes speech from an audio or video and converts it into text. \par

Different methods for STT and text-to-speech (TTS) have been reviewed in \cite{Nagdewani2020}. The basic process discussed in the paper for STT involves feature extraction, word matching using acoustic word models, sentence matching using syntax and semantics and finally language modelling to text. \par

Recognizing emotions from text is often classed as Sentiment Analysis (SA), however, if the textual classification is done on more than just classifying it into positive, negative and neutral, could provide more depth and context to the text. In \cite{Acheampong2020}, three ways to approach text-based emotion detection as rule construction, ML and hybrid where rule-based approach uses the grammatical and logical rules to find emotions from a document, the ML approach uses ML algorithms to classify text into emotions and hybrid approach is the combination of both are presented.  \par

In recent years, with advances in the DL methods, most of the research has been done using convolutional neural networks in a varied field of applications. A fine-tuned VGGNet architecture, in \cite{Khaireddin2021}, based on CNN is used to achieve the state-of-the-art single network precision of 73. 28\% on the FER-2013 dataset. \cite{Teja2023}'s researcher uses SVM, CNN and pre-trained VGG-16 models to recognize emotions from facial expression along with age and gender prediction. The author of \cite{Pomazan2023} explores the potential of CNN-based emotion recognition in marketing research for advertising and gains insight into consumer behavior. In \cite{Zbaida2023}, CNN-based technology is incorporated into e-learning platforms by comparing five different algorithms (VGG16, VGG19, RESNET50V2, EfficientNETB0 and EfficientB7) on their accuracy to find the best-suited algorithm. However, among all these CNN applications, a recent trend of Vision Transformers (ViT) has emerged for computer vision tasks. \par

 The potential of ViT-based models is explored in \cite{Bobojanov2023} for FER on three datasets; RAF-DB, FER2013 and a clean, augmented and balanced dataset using images from the FER dataset. They conducted a comprehensive evaluation of 13 ViT models on these datasets and concluded a promising success of these models for FER tasks. A ViT method introduced in \cite{Huang2021}, to recognize driver expressions by performing data augmentation based on parallel imaging framework with the StarGAN network on CK + and KMU-FED data sets, achieved a higher recognition rate than CNN and ResNet18. \par
 
Following the trend and being inspired by \cite{Huang2021}, this paper explores a different ViT based model in the application of user testing to enhance the user experience. Text-based emotion detection would provide more contextual understanding by providing context to emotions. A combination of STT and Text-based emotion detection would provide deeper and nuanced emotional insights than directly using the SER which only focuses on the tone of the speaker to recognize emotions.

\subsection{Emotion Recognition in User Testing}
While reviewing different applications of emotion recognition, it was observed that little has been done in the field of user testing. Emotion recognition would not only provide deeper meaningful insights by capturing verbal and non-verbal emotions but also help in improving design decisions leading to better product performance. Limited literature was found in which the problem of manually keeping track of user emotions, be it facial or speech, has been addressed. The author in \cite{Veriscimo2021} also agrees with the fact that little has been done in the evaluation of UX through automated emotion recognition. The author presents a systematic review on implementation of FER in the UX evaluation thus concluding the need to attend the lacking standardization and modernization of tools, procedures, and evaluation criteria for emotion recognition in UX. \par

Some efforts have been made to decrease the gap. In \cite{Razzaq2023}, a hybrid multimodal emotion recognition framework was introduced for UX evaluation, achieving an average accuracy of 98. 19\% on detecting four emotions; happiness, neutral, sadness, and anger. A UX evaluation model using an automated tool to calculate the user experience rating of a digital product / service in positive, negative, and neutral points called UXAPP has been implemented and validated in \cite{Cordeiro2024}. The author used state-of-the-art emotion recognition modalities to carry out implementation and conducted experiments to generate a report through nine individuals who carried out the designated task and finally compared the results with the generated report of UXAPP. \par

Even after implementing different modalities to automate the process of manually keeping track of user emotions, it would not make sense for UX designers to handle gathered data manually.  In order to automate the data analysis/insight generation process  and being inspired by \cite{Cordeiro2024}, rather than generating a positive, negative and neutral point system, a way to make the user ask questions directly from the gathered data rather than manually analyzing was explored in this project. \par

The concept of a whole process from evaluating the video to gathering insights and analyzing the gathered data through automation has been demonstrated using this project.

\section{Conceptual Framework}
\begin{figure}[htbp]
    \centering
    \includegraphics[width=\textwidth]{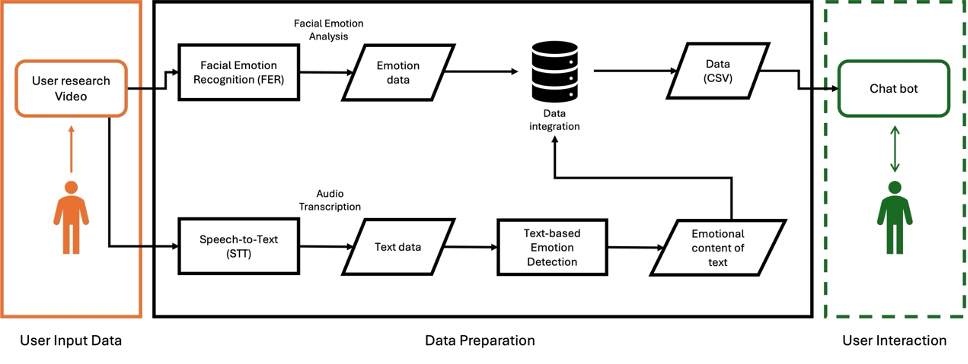} 
    \captionsetup{justification=centering}
    \caption{Project architecture.}
    \label{fig:fg4}         
\end{figure}

The following architecture has been devised to address the stated problem (Figure ~\ref{fig:fg4}). The architecture was divided into 3 stages:
\begin{itemize}
    \item User research input video
    \item Data preparation
    \item User interaction 
\end{itemize}

\begin{description}
    \item[Stage 1] The UX designer or researcher provides a pre-recorded video session of usability testing to the framework.
    \item[Stage 2] In the data preparation phase, the input video is separately fed into a facial emotion recognition (FER) model and a speech-to-text (STT) model. The FER model detects the face in the video frame-by-frame and recognizes emotions expressed by the person in the video. The emotions detected are recorded and presented as probability in each frame. The data gathered is saved in a CSV file. The STT model converts the spoken language into written text i.e. audio transcription while recording the start and end time of each statement of the speaker. The output text is then fed to the Text-based emotion detection model which classifies the text into different emotions described in the model used for classification with the confidence score of the detected emotion. The final output is then saved in a separate CSV file.  \par
The data from both files are then integrated into a final data file. This is further fed to the AI chatbot through the conversational analyst tool for user interaction with the data. 
    \item[Stage 3] The user can interact with the chatbot for any queries related to the video provided, reducing the manual labor of re-watching the video for emotion tracking. \par
The integration of facial and speech expression data would also enhance the chances of finding discrepancies and anomalies in the video to get in-depth insights. 
\end{description}

\subsection{Facial Emotion Recognition}

\begin{figure}[htbp]
    \centering
    \includegraphics[scale =0.8]{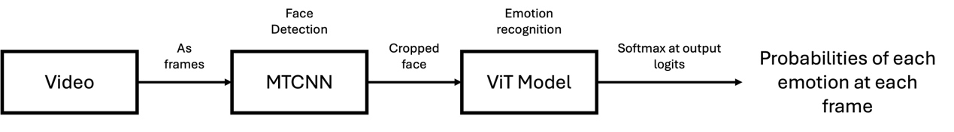}
    \captionsetup{justification = centering}
    \caption{FER model workflow}
    \label{fig:fg5}
\end{figure}

Facial emotion recognition, or FER, is used to detect and recognize faces in an image or video and classify them into basic emotions that give in depth insight into the person's feelings through facial expressions. The authors in \cite{Setiowati2017}, through experiments and comparisons, demonstrates the higher accuracy of DL methods. In this project, the DL methods were implemented to gather insights from user research video. MTCNN network was used to detect faces and the Vision Transformer (ViT) model was used for facial emotion recognition. The out-performance of the MTCNN architecture from other state-of-the-art CNN methods was demonstrated in \cite{Zhang2016}, and according to \cite{Dosovitskiy2020}, ViT is relatively inexpensive to pre-train and matches or exceeds the state-of-the-art on many image classification datasets. Thus, leading to the decision of using these method and model. The general workflow for facial emotion recognition can be seen in Figure ~\ref{fig:fg5}. \par

\subsection{Speech-to-Text and Text-based Emotion Detection}
A machine learning technology that allows computers to understand, analyze, manipulate, interpret, and comprehend human language is known as Natural Language Processing (NLP). In day-to-day life corporations and business gather a lot of textual data such as emails, surveys, and speech data such as voice calls which are needed to be processed and analyzed to understand the intent or sentiment in the message and respond in real time to human communication. NLP software makes it easy to handle this more conveniently and efficiently. The technologies used in this phase of the project fall under NLP tasks. \par

Speech-to-Text is a technology used to recognize and transcribe spoken language in an audio, video, or from a dictation to text through computational linguistics. It is also known as Automatic Speech Recognition (ASR). Its applications range from home applications to industrial such as media, banking, and medical.\par

\begin{figure}[H]
    \centering
    \includegraphics[scale = 0.8]{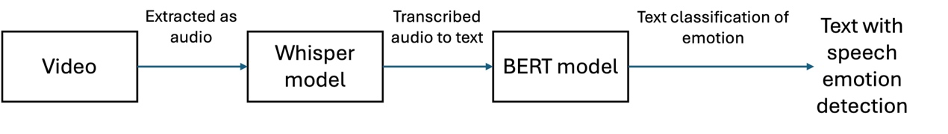}
    \captionsetup{justification = centering}
    \caption{Overall framework of STT and Text-based Emotion detection.}
    \label{fig:fg6}
\end{figure}

Text-based emotion detection involves identifying and analyzing the underlying emotions in a text. It can also be categorized as sentiment analysis where the main focus is unraveling any kind of emotions or sentiments expressed in a given text. Most of its applications are seen in the fields of healthcare, customer support, marketing, and HCI. In this project, both technologies are used to unravel the underlying emotions expressed by the speech of the participant while reviewing a product or web design during user testing. The overall framework is illustrated in Figure ~\ref{fig:fg6}.

For ASR, the Whisper model would be implemented. This technology will help recognize the speech of the participant and transcribe it. In \cite{Vásquez2023}, the author compared the use of two prominent technologies for ASR at the moment, namely Wav2vec2.0 and Whisper, for forensic scenarios, and concluded that the latter was more accurate in most cases, especially in uncontrolled acoustic conditions. The researchers evaluated both models in corpora of seven Indo-European languages, additionally inventing their own in-domain corpus using different open source datasets from the community research with an aim of testing the models in more realistic and operative conditions. Considering the good results and dynamic capabilities of Whisper, it was chosen for ASR implementation. \par 

The researchers in \cite{Radford2022} describe the architecture and flow of Whisper. The model works on sequence-to-sequence Transformer-based architecture. The model converts the raw audio input into log-Mel spectrograms which are the 2D representations of audio, where one dimension represents time and other represents frequency bands, before passing it to the Encoder. A Mel spectrogram represents the audio in the frequency domain. As the transformer doesn’t know about the order of the input sequence, a Sinusoidal Positional Encoding is added to the input before it is fed to the Encoder. \par

The BERT (Bidirectional Encoder Representations from Transformers) model was introduced by \cite{Devlin2019} to pre-train deep bidirectional representations from unlabeled text. As the name of the model suggests, its architecture is Transformer-based, specifically using only the Encoder part. The model has a bidirectional nature, meaning that rather than only reading the text from left-to-right or right-to-left, it reads the sequence of words at once considering the context from both directions. In \cite{Devlin2019}, the model is developed based on the original implementation by \cite{Vaswani2017}. A detailed description of the model is provided in \cite{Devlin2019}.

\subsection{Data Integration and Insight Generation}
Data integration is a crucial step in this project. Analyzing integrated data from different modalities such as facial expressions and speech emotion detection would allow to have a comprehensive and insightful analysis of user research videos. A single model may not be sufficient to grasp the emotions of a person. As discussed in emotion theories and models, emotions can be expressed in different ways such as voice, body language, face, etc., and a combination of different modalities would help to capture the whole picture of emotions expressed in the video. Other benefit of joining the data is that it would help find any anomalies or patterns such as instances where the facial expressions is happy, but the speech suggests sarcasm or say otherwise, or instances where they align perfectly. This would provide more accurate interpretation and reliability of the gathered data. \par

Sometimes data generated in one modality is prone to errors, missing or misleading data, and the integration process would help to mitigate these errors by providing additional context and validation. For example, while detecting emotions from one of the user research videos (which was later not used in experiments), the FER model was not able to detect any face and reported 80\% of the data as empty, that is, No face detected, while speech emotion detection went smoothly and classified the data into emotions. In such scenarios, even though little face data were available, the speech data helped in retaining the emotions expressed by the participant, which was better than nothing. \par

In addition, data integration would help analyze emotions in context, leading to more meaningful insights. For example, a combination of sad expression with a negative speech tone would suggest a deeper level of dissatisfaction from the participant. \par

In terms of ethics and responsibility, the combination of different data sources would respect the complexity of human emotions by reducing misinterpretations, and from a practical perspective, data integration would reduce the complexity of interpreting and handling multiple streams of data separately, allowing more streamlined data processing. In this project, data integration was implemented keeping in mind the benefits it offers. \par

For data integration, two concepts of Temporal alignment and Temporal fusion are used. Temporal alignment refers to synchronizing data from different data sources based on their time-related attributes, whereas Temporal Fusion refers to combining different time-aligned data from different sources into a single representation to obtain a more comprehensive view of the data.\par 

With the help of AI, it is now possible to interact with the data directly without the need to learn SQL queries or hard coding, making the data analysis process automated. In this project, one of the possible solutions, a Python library called PandasAI, was used, which allows users to interact with the data in a natural language way, also known as a conversational data analyst tool. The link can be found at \cite{Sinaptik-AI2024}.\par

The tool allows to use different Large Language Models (LLMs) such as default BambooLLM, OpenAI models, Google PaLM, Google Vertexai, Azure OpenAI, HuggingFace via Text Generation, Amazon Bedrock models, IBM watsonx.ai models, and local models such as Ollama and LM Studio. Incorporation of such technologies in the application of User Testing would allow designers to evaluate the user research videos in less time and in more efficient way. 

\section{Case Study}
\subsection{Facial Emotion Recognition}
In this study, a pre-trained ViT model was used for facial emotion recognition. In order to choose a suitable model from 10 pre-trained working ViT models available on website called HuggingFace, the models were evaluated on the three benchmark datasets i.e. FER-2013, AffectNet and CK+48.\par
The FER-2013 dataset contains 28709 training and 5404 testing 48x48 pixels grayscale images of 7 emotions namely angry, disgust, fear, happy, sad, surprise and neutral. The AffectNet dataset contains high dimensional coloured 37553 training, 3200 testing and 800 validation images of 8 emotions namely angry, disgust, fear, sadness, happy, surprise, contempt and neutral. The CK+48 dataset contains 981 (48x48 pixels grayscale) images of 7 emotions namely angry, disgust, happy, sadness, fear, surprise and contempt. 

The model selection was done by evaluating the available models on the three datasets on the basis of accuracy, F1-score, precision and recall. The performance of models on different datasets can be seen in Table ~\ref{tab:tb1} below:  

\begin{table}[H]
    \centering
    \resizebox{\textwidth}{!}{  
        \begin{tabular}{|c|c|c|c|c|c|c|c|c|c|c|c|c|}
            \hline
            \multirow{3}{*}{\textbf{Model No}} & \multicolumn{12}{|c|}{\textbf{Performance Evaluation on Datasets}} \\ \cline{2-13}
            & \multicolumn{4}{|c|}{\textbf{FER-2013}} & \multicolumn{4}{|c|}{\textbf{CK+48}} & \multicolumn{4}{|c|}{\textbf{AffectNet}} \\ \cline{2-13}
            & Accuracy & Precision & Recall & F1-Score & Accuracy & Precision & Recall & F1-Score & Accuracy & Precision & Recall & F1-Score \\ \hline
            1  & 0.7115 & 0.7076 & 0.6935 & 0.6997 & 0.3649 & 0.2648 & 0.3060 & 0.2802 & 0.2675 & 0.3177 & 0.2675 & 0.2233 \\ \hline
            2  & 0.3536 & 0.3433 & 0.3094 & 0.2751 & 0.2120 & 0.3116 & 0.2460 & 0.1991 & 0.2128 & 0.1594 & 0.2128 & 0.1604 \\ \hline
            3  & 0.0906 & 0.6176 & 0.1417 & 0.0619 & 0.0479 & 0.5827 & 0.1164 & 0.0199 & 0.1019 & 0.0418 & 0.1019 & 0.0542 \\ \hline
            4  & 0.1106 & 0.1308 & 0.2222 & 0.0945 & 0.3191 & 0.3331 & 0.4253 & 0.2347 & 0.1297 & 0.0980 & 0.1297 & 0.1055 \\ \hline
            5  & 0.1209 & 0.5090 & 0.1346 & 0.0977 & 0.1458 & 0.5146 & 0.1876 & 0.1162 & 0.1316 & 0.0620 & 0.1316 & 0.0795 \\ \hline
            6  & 0.1123 & 0.1124 & 0.2244 & 0.0772 & 0.4139 & 0.4155 & 0.4239 & 0.2776 & 0.1263 & 0.0839 & 0.1263 & 0.0910 \\ \hline
            7  & 0.0741 & 0.1318 & 0.1961 & 0.0650 & 0.3848 & 0.4488 & 0.4274 & 0.2629 & 0.1097 & 0.0872 & 0.1097 & 0.0954 \\ \hline
            8  & 0.0911 & 0.1203 & 0.2002 & 0.0723 & 0.1804 & 0.1682 & 0.2523 & 0.1300 & 0.1013 & 0.1027 & 0.1013 & 0.0918 \\ \hline
            9  & 0.1180 & 0.1074 & 0.2203 & 0.0876 & 0.3170 & 0.3058 & 0.3959 & 0.2549 & 0.1141 & 0.0995 & 0.1141 & 0.1037 \\ \hline
            10 & 0.0984 & 0.1182 & 0.0751 & 0.0733 & 0.4098 & 0.3601 & 0.4354 & 0.2859 & 0.1062 & 0.0955 & 0.1062 & 0.0983 \\ \hline
        \end{tabular}
    }
    \caption{Performance evaluation of models on different datasets. Model names are as follows: 1. trpakov/vit-face-expression, 2. jayanta/vit-base-patch16-224-in21k-emotion-detection, 3. Hector001/emotion-vit-model-hector, 4. hilmiatha/ViT-Emotion-Classifier, 5. StoneSeller/emotion-classifier-vit, 6. andikamandalaa/vit-base-patch16-224-in21k-emotion-classification, 7. gabrielganan/vit-emotion\_classification, 8. andikamandalaa/vit-base-patch16-224-in21k-emotion-classification3, 9. evanrsl/vit\_facial\_emotion, 10. asyafalni/vit-emotion-classifier.}
    \label{tab:tb1}
\end{table}
Observing the performances on different datasets, Model 1 (trpakov/vit-face-expression) was chosen for Facial Emotion Recognition process proving to have more generalization ability and therefore can be used in different scenarios. The specifications of the model can be found at \cite{Trapkov2024}.\par

The user testing videos for the demonstration of the concept were chosen from YouTube. The 2 chosen videos were labeled as Testing\_video1 and Testing\_video2 for anonymity. Usability testing videos are usually large and, due to hardware limitations, the frame rate of video file was decreased from 30 fps to 15 fps for Testing\_video1.mp4 along with some video quality compressions. The Testing\_video1 was also cut short to 9 mins and 32 seconds, removing the introduction and 1 task performed by the participant, from 14 mins and 39 seconds video. \par

The input video is fed as video frames to the MTCNN architecture. MTCNN group model detects the face and if a face is found, coordinates of the bounding box drawn on the face are recorded, and the face is cropped from the video frame (each frame can be seen as an image). This cropped face is preprocessed using an auto-feature extractor and fed to the pre-trained ViT model. The model recognizes the faces in each frame and classifies them as basic emotions as defined in the model configuration. To get the probabilities of each emotion in each frame, a softmax loss function is applied to the output logits. ‘Id2label’ attributes of the pre-trained ViT model are retrieved from the configuration and class labels (emotions defined in the model) are mapped to their respective probabilities. The output data generated is then saved in a CSV file for later use.\par

A function was created taking face image and emotion probabilities as input to generate combined images consisting of two parts as face image and emotion probabilities bar plot, with the original face on the left and the bar plot on the right. 




The total duration of the generated video was 9 min and 16 sec. All data generated by facial emotion recognition was stored in a CSV file called 'emotion\_probabilities.csv'. \par 

\begin{figure}[H]
    \centering
    \includegraphics[width=\linewidth]{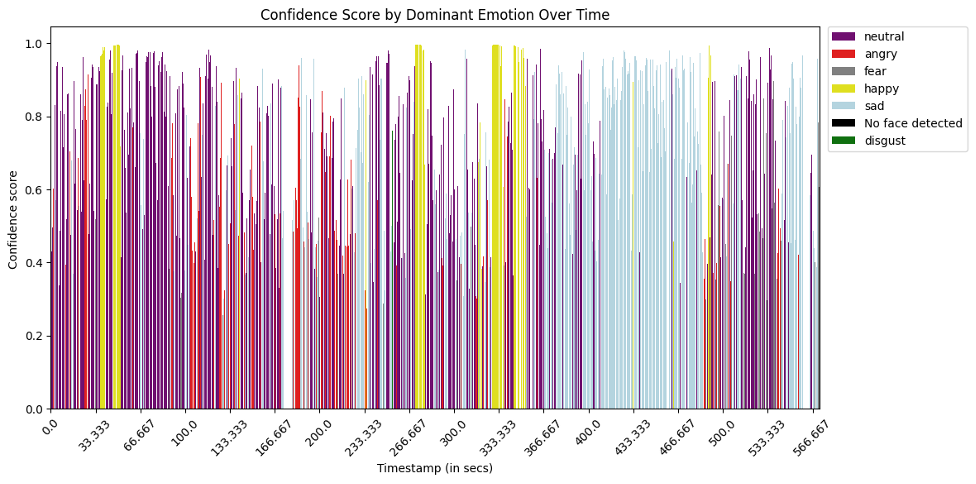}
    \captionsetup{justification=centering}
    \caption{Normalized values of all emotions detected at each frame in Testing\_video1}
    \label{fig:fg11}
\end{figure}

In Figure \ref{fig:fg11}, it can be seen that the participant mainly felt ‘sad’ for a duration of 100 seconds (400-500 secs in the graph). There are some instances of ‘happy’ and ‘angry’ emotions but overall, the participant had a ‘neutral’ face while performing the assigned tasks. These little changes in the emotions while reviewing provide an important information to the UX designers as they can then check at what time the participant was happy or unhappy while performing tasks.

\subsection{Speech-to-Text and Text-based Emotion Detection}




In this implementation phase, the video was loaded using the 'VideoFileClip()' function of the 'moviepy.editor' package. The audio was extracted and written in the '.wav' extension. After extracting the audio, the speech recognizer was initialized. The base version of the whisper model was loaded transcribing the recognized speech in the audio file. The balance between performance and computational efficiency is provided by the ‘base’ version of the model. Since the resultant text was timestamped word-by-word and segmented, it was concatenated to form the complete transcribed text and then tokenized into sentences. This step was necessary for later analysis in the Data Integration part. \par

A pre trained classifier and tokenizer model called ‘bhadresh-savani/bert-base-go-emotion’ from HuggingFace was used for text classification into emotions. In \cite{Bhadresh2024}, the authors evaluated the model attaining an accuracy of 96.14\% with 0.116 evaluation loss and 0.12 as training loss. The model was trained on 169208 instances, using 3 epochs with 16 as batch size and 31728 optimization steps. A function was created that took the segments and the classifier as input and classified each segment of text within the list of segments using the given classifier. The results included a list of dictionaries in which each dictionary contained the original text, start and end time of the sentence, emotion class of the text, and its confidence score. Table \ref{tab:tb2} illustrates the resultant classification of recognized text from audio into emotions. The resultant data was saved in a CSV file named ‘speech\_probabilities.csv’. 

\begin{table}[htbp]
\centering
\begin{tabularx}{\textwidth}{|X|c|c|c|c|}
\hline
\textbf{Text} & \textbf{Start} & \textbf{End} & \textbf{Emotion} & \textbf{Confidence} \\ \hline
We'll be going to be posting that next part of the task. & 166.92 & 170.52 & neutral & 0.3332 \\ \hline
And I'm just going to kind of get us back to the homepage so we can restart. & 171.8 & 176.72 & neutral & 0.2554 \\ \hline
Okay. & 179.0 & 179.22 & caring & 0.4461 \\ \hline
So next task. & 179.8 & 181.2 & neutral & 0.9552 \\ \hline
You're going camping this weekend but you don't have a tent. & 182.34 & 185.62 & neutral & 0.2950 \\ \hline
You want to find and rent a two-person tent to use. & 186.02 & 189.0 & neutral & 0.7209 \\ \hline
Use Surfboard Board to accomplish this. & 189.52 & 191.42 & neutral & 0.8308 \\ \hline
Okay. & 191.42 & 192.68 & caring & 0.4461 \\ \hline
So I'm going to get here this time. & 193.62 & 196.42 & neutral & 0.3660 \\ \hline
And I'm looking for a tent. & 197.32 & 199.64 & neutral & 0.7322 \\ \hline
I want a two-person tent. & 201.74 & 205.16 & desire & 0.4894 \\ \hline
I'm not really sure if I should click the menu looking thing or take an icon. & 207.0 & 214.26 & confusion & 0.4445 \\ \hline
\end{tabularx}
\captionsetup{justification = centering}
\caption{Text classification into emotions from recognized speech of Testing\_video1.}
\label{tab:tb2}
\end{table}



\begin{figure}[H]
    \centering
    \includegraphics[width=\linewidth]{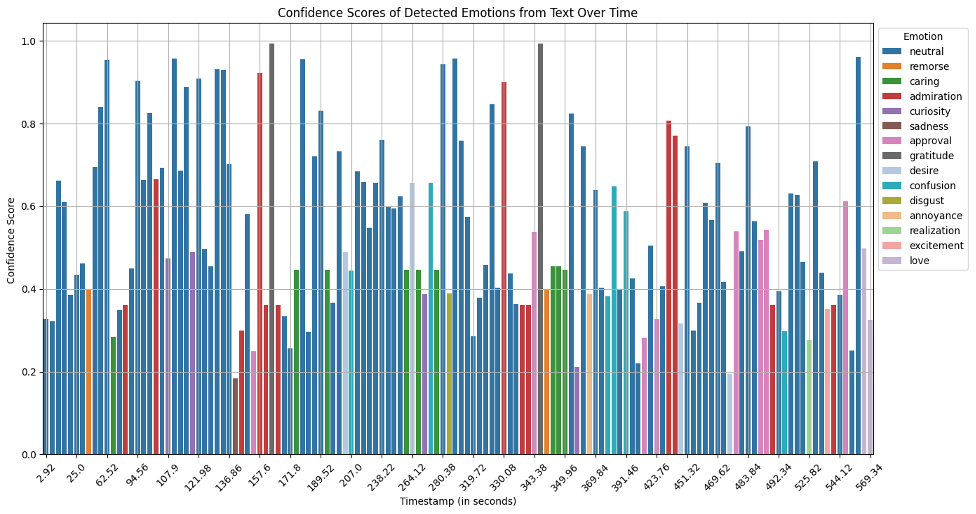}
    \captionsetup{justification=centering}
    \caption{Emotions detected from participant's speech over time from Testing/\_video1}
    \label{fig:fg15}
\end{figure}

Figure \ref{fig:fg15} illustrates that the participant’s statement had different emotions over the course of the video. Between 369 and 391 seconds, the emotion of 'confusion' was quite dominant, indicating the participant's discomfort in completing the task during that period. Comparing the results from facial emotion recognition, between these time periods, the user also had sad expressions, thus providing deeper insights.

\subsection{Data Integration and Insight Generation}

Two types of data have been collected:
\begin{enumerate}
    \item Facial Emotion data
    \item Speech Emotion data
\end{enumerate}
\textbf{Facial emotion data} was collected from emotions detected by facial expression of the participant. The data was saved in a CSV file called ‘emotion\_probabilities.csv’. \par

\textbf{Speech Emotion Data} was gathered using the combination of STT and Text-based emotion detection. The ‘speech\_probabilities.csv’ file was created to save the data for speech emotion detection. There were no modifications done as the file was clean and ready to use. The file contains five columns, namely ‘text’ – containing the participant’s statements, ‘start’ – the start time (in seconds) of statement made by the participant, ‘end’ – the end time (in seconds) of statement made by the participant, ‘emotion’ – the detected emotion of the text and ‘confidence’ – the confidence score of the detected emotion. \par

\textbf{Data cleaning and preparation}:
There were some instances where the model could not detect any face and returned NULL values. To handle these, they were converted to numeric 0 and not deleted as they were considered as important information over the time period. To obtain the dominant facial emotion in each frame, the emotion with the highest probability score was used. For data without a record, the dominant emotion was converted to the value 'No face detected'. After all transformations, the data was saved in another CSV file called ‘non\_zero\_dfFER.csv’. \par

The data file contains three columns, that is, Timestamp – time (in seconds) at which the emotion was recorded, ‘Highest Score’ – the confidence score of the detected emotion and ‘Facial Emotion’ – the dominant emotion at the time of detection.\par

After data collection, cleaning, and preparations, the ‘non\_zero\_dfFER.csv’ and ‘speech\_probabilities.csv’ files were used for data integration. Implementation was done in two steps, i.e. Temporal alignment and Temporal Fusion. \par

\textbf{Temporal alignment} was implemented in the implementation stages of collecting facial emotion data and speech emotion data. The ‘Timestamp’ variable was added to the facial emotion data and the ‘start’ and ‘end’ variable in the speech emotion data. To make sure that the data was aligned according to the time, the variables were converted to the H-M-S (hours-minutes-seconds) format from the using the datetime() function from the Pandas library before integration. This step was taken to accurately compare and combine the two different data sources.\par 

\textbf{Temporal Fusion}  was done by aggregating or fusing the facial emotion data based on time intervals of the speech emotion data. Simply explaining, if the participant’s statement initiated at 2 second of recorded time and ended at 4 seconds, the number of frames present in that time interval were aggregated to represent the most frequently detected emotion within that same interval of time. \par

In addition, the fusion was done considering two scenarios. For the number of frames in the specified time interval, if the emotions detected in all the frames are same, then the average confidence score for facial emotion data at that time would be ‘mean’ of all the data points representing the same emotion. Secondly, if the emotions detected in the facial emotion data are not same within the specified time interval, the most frequent emotion is extracted and the average of confidence score for only those particular frames would be considered as the final score for the detected emotion. For example, if within time interval of 2 seconds, 30 frames were captured, and for all 30 frames the emotion detected was ‘neutral’, then the Facial Emotion for that time would be ‘neutral’ with the confidence score as the average of Highest Score of all 30 frames. And if among 30 frames, 20 are detected as ‘happy’ and 10 as ‘neutral’, the Facial Emotion for that time would be ‘happy’ as it is the most frequent one with confidence score as the average of Highest Score of only those 20 frames. \par

Furthermore, if the facial emotion data were not in alignment with the speech emotion data, logic would return the 'None' value. \par

The implementation was done in Python version 3.12.3 in VS Code. Temporal alignment was performed using the datetime() function. A ‘for’ loop was initiated by iterating through each row of the speech emotion data. After finding the facial emotion data that fall within the current time interval of the speech emotion data, temporal fusion was implemented. The resultant integrated data was saved in ‘Test1.csv’ file. A snippet of the final integrated data file is shown in Table \ref{tab:tb5}. \par

\begin{table}[H]
\centering
\resizebox{\textwidth}{!}{
\begin{tabular}{|l|c|c|c|c|c|c|}
\hline
\textbf{Text} & \textbf{Start} & \textbf{End} & \textbf{Emotion} & \textbf{Confidence} & \textbf{Avg FER Score} & \textbf{Dominant FER Emotion} \\ \hline

So I'm going to print it out. & 2.92 & 5.96 & Neutral & 0.326197386 & 0.815838384 & Neutral \\ \hline
So I'm going to click List Your Gear. & 6.44 & 9.24 & Neutral & 0.321020454 & 0.834265302 & Neutral \\ \hline
And that Own Attent. & 13.82 & 15.86 & Neutral & 0.661488712 & 0.736815128 & Angry \\ \hline
So I'm going to click Tent. & 16.3 & 17.94 & Neutral & 0.61031723 & 0.745623811 & Neutral \\ \hline
Give your listing a descriptive title. & 22 & 25 & Neutral & 0.385479093 & 0.807047625 & Neutral \\ \hline
And I'm going to click the check box listed. & 25 & 32.22 & Neutral & 0.433583885 & 0.77720179 & Neutral \\ \hline
Yeah, this is the Tent. & 47 & 48.3 & Neutral & 0.461419493 & 0.994635472 & Happy \\ \hline
Sorry. & 49 & 49.2 & Remorse & 0.399663627 & 0.995449468 & Happy \\ \hline
Now I'm going to write a description about my Tent. & 49.2 & 55.12 & Neutral & 0.69373101 & 0.955230098 & Happy \\ \hline
So this here. & 56.5 & 59.78 & Neutral & 0.839199185 & 0.809589095 & Neutral \\ \hline
And this is about three people. & 62.52 & 68.36 & Neutral & 0.952671409 & 0.870423921 & Neutral \\ \hline
Two, and if you move a lot in your sleep, it's pretty light. & 68.36 & 82.34 & Caring & 0.28432548 & 0.86390207 & Neutral \\ \hline
Okay, I'm going to type that correctly. & 84.12 & 87.24 & Neutral & 0.349999487 & 0.783816478 & Neutral \\ \hline
Cool. & 88.86 & 89.98 & Admiration & 0.361270905 & 0.577821005 & Neutral \\ \hline
And I'm going to set my price. & 90.32 & 92.08 & Neutral & 0.450031757 & 0.580378833 & Angry \\ \hline
Let's see. & 94.56 & 95.42 & Neutral & 0.903746367 & 0.521529078 & Neutral \\ \hline
Probably just put it up for like \$5 per day. & 97.06 & 101.04 & Neutral & 0.663080513 & 0.725880786 & Neutral \\ \hline
It's like \$100 Tent. & 101.42 & 102.7 & Neutral & 0.824869871 & 0.676134035 & Sad \\ \hline
That would be pretty cool. & 103.1 & 104 & Admiration & 0.665003479 & 0.732171007 & Neutral \\ \hline
I'll pick up Address. & 105.36 & 106.62 & Neutral & 0.692145646 & 0.714218024 & Sad \\ \hline
Where will people pick it up from you? & 107.9 & 109.84 & Curiosity & 0.474241048 & 0.640520512 & Angry \\ \hline
And you see a C. & 112.06 & 114.12 & Neutral & 0.956321597 & 0.821540532 & Neutral \\ \hline
And take a look at this. & 115.4 & 117.64 & Neutral & 0.684900999 & 0.881912405 & Neutral \\ \hline
Like 10. & 117.78 & 118.28 & Neutral & 0.887729764 & 0.955703162 & Neutral \\ \hline
What's this? & 118.54 & 119 & Curiosity & 0.488924116 & 0.960007565 & Neutral \\ \hline
And then we've got my product details. & 121.98 & 124.76 & Neutral & 0.90899092 & 0.666946242 & Neutral \\ \hline
I'm going to click here. & 125.34 & 126.4 & Neutral & 0.49662587 & 0.53009551 & Angry \\ \hline
\end{tabular}
}
\captionsetup{justification = centering}
\caption{The resultant data integrated "Test1.csv" file.}
\label{tab:tb5}
\end{table}

\subsubsection{Insight Generation}
Among all the available options, Open AI's GPT- 4o was used to evaluate the data collected by asking queries from the 'Test1.csv' file. In the query session with the LLM, the following questions were asked. 

\begin{enumerate}
    \item What was the most dominant facial emotion and how many times was it detected?
    \item At what time did the user feel the most happy? Can you provide the statement and the confidence score?
    \item How many times has the user felt confusion/frustration in the video, and can you list the corresponding data?
    \item Was there a time where no face was detected in the video?
    \item Can you find anomalies in speech and facial emotions?
    \item If I were to map the speech emotions into facial emotions, were there any anomalies where speech emotions did not match facial emotions?
    \item Can you find where the user recommended any improvement or would like to have something different from what was presented?
    \item This data is based on user testing, as part of user experience study, can you tell me the key places where the experience can be improved?
\end{enumerate}
\par
To get an overview of the emotions expressed by the participant/user, there is always a question about the dominant emotion detected. When asked the same question, it responded as shown in Table \ref{tab:tb6} Q1. The ‘neutral’ emotion detected as the dominant facial emotion also aligns with previous findings that show a positive result of using the LLM as a conversational data analyst tool. \par

\begin{table}[htbp]
  \centering
  \resizebox{\textwidth}{!}{
    \begin{tabular}{c|l|p{13.75em}r|ll|l|l|l|l|}
    \toprule
    \rowcolor[rgb]{ .851,  .851,  .851} \multicolumn{1}{|c|}{\textbf{Number}} & \multicolumn{9}{c|}{\textbf{Question \& Answers}} \\
    \midrule
    \rowcolor[rgb]{ .949,  .949,  .949} \multicolumn{1}{|c|}{\textbf{1}} & \multicolumn{9}{l|}{\textbf{What was the most dominant facial emotion and how many times was it detected?}} \\
    \midrule
          & \multicolumn{9}{l|}{{'type': 'string', 'value': "The most dominant facial emotion is 'neutral' and it was detected 63 times."}} \\
\cmidrule{2-10}    \multicolumn{1}{c}{} & \multicolumn{1}{l}{} & \multicolumn{1}{l}{} & \multicolumn{1}{l}{} &       & \multicolumn{1}{l}{} & \multicolumn{1}{l}{} & \multicolumn{1}{l}{} & \multicolumn{1}{l}{} & \multicolumn{1}{r}{} \\
    \midrule
    \rowcolor[rgb]{ .949,  .949,  .949} \multicolumn{1}{|c|}{\textbf{2}} & \multicolumn{9}{l|}{\textbf{At what time did the user feel the most happy? Can you provide the statement and the confidence score?}} \\
    \midrule
          & \multicolumn{9}{p{71.745em}|}{{'type': 'string', 'value': "The user showed the most happy face at 330.08 seconds with the statement: ' Returning to the one I posted, it sounds pretty awesome.' and confidence score: 0.899553418."}} \\
\cmidrule{2-10}    \multicolumn{1}{r}{} & \multicolumn{1}{r}{} & \multicolumn{1}{r}{} & \multicolumn{1}{r}{} &       & \multicolumn{1}{r}{} & \multicolumn{1}{r}{} & \multicolumn{1}{r}{} & \multicolumn{1}{r}{} & \multicolumn{1}{r}{} \\
    \midrule
    \rowcolor[rgb]{ .949,  .949,  .949} \multicolumn{1}{|c|}{\textbf{3}} & \multicolumn{9}{l|}{\textbf{How many times has the user felt confusion/frustration in the video, and can you list the corresponding data?}} \\
    \midrule
          & \textbf{index} & \multicolumn{2}{p{20.915em}|}{\textbf{text}} & \multicolumn{1}{l|}{\textbf{start}} & \textbf{end} & \textbf{emotion} & \textbf{confidence} & \textbf{avg\_fer\_score} & \textbf{dominant\_fer\_emotion} \\
\cmidrule{2-10}          & 50    & \multicolumn{2}{p{20.915em}|}{I'm not really sure if I should click the menu looking thing or take a icon.} & \multicolumn{1}{l|}{207} & 214.26 & confusion & 0.444 & 0.582 & neutral \\
\cmidrule{2-10}          & 63    & \multicolumn{2}{p{20.915em}|}{I'm not really sure how to find.} & \multicolumn{1}{l|}{277.82} & 279.86 & confusion & 0.655 & 0.497 & neutral \\
\cmidrule{2-10}          & 92    & \multicolumn{2}{p{20.915em}|}{I clicked it and then I wasn't really sure exactly how to filter three results.} & \multicolumn{1}{l|}{376.54} & 382.46 & confusion & 0.381 & 0.785 & sad \\
\cmidrule{2-10}          & 93    & \multicolumn{2}{p{20.915em}|}{I'm not sure what that was.} & \multicolumn{1}{l|}{384.84} & 387.62 & confusion & 0.648 & 0.633 & sad \\
\cmidrule{2-10}          & 95    & \multicolumn{2}{p{20.915em}|}{I'm not really sure these three I can't do.} & \multicolumn{1}{l|}{391.46} & 393.54 & confusion & 0.587 & 0.725 & neutral \\
\cmidrule{2-10}          & 121   & \multicolumn{2}{p{20.915em}|}{" So I'm never used Airbnb or anything, but I guess the most similar product I have used"} & \multicolumn{1}{l|}{499.8} & 507.4 & confusion & 0.296 & 0.727 & neutral \\
\cmidrule{2-10}    \multicolumn{1}{c}{} & \multicolumn{1}{l}{} & \multicolumn{1}{l}{} & \multicolumn{1}{l}{} &       & \multicolumn{1}{l}{} & \multicolumn{1}{l}{} & \multicolumn{1}{l}{} & \multicolumn{1}{l}{} & \multicolumn{1}{r}{} \\
    \midrule
    \rowcolor[rgb]{ .949,  .949,  .949} \multicolumn{1}{|c|}{\textbf{4}} & \multicolumn{9}{l|}{\textbf{Was there a time where no face was detected in the video?}} \\
    \midrule
          & \textbf{index} & \multicolumn{2}{p{20.915em}|}{\textbf{text}} & \multicolumn{1}{l|}{\textbf{start}} & \textbf{end} & \textbf{emotion} & \textbf{confidence} & \textbf{avg\_fer\_score} & \textbf{dominant\_fer\_emotion} \\
\cmidrule{2-10}          & 40    & \multicolumn{2}{p{20.915em}|}{And I'm just going to kind of get us back to the homepage so we can restart.} & \multicolumn{1}{l|}{171.8} & 176.72 & neutral & 0.255 & 0     & No face detected \\
\cmidrule{2-10}          & 84    & \multicolumn{2}{p{20.915em}|}{No worries.} & \multicolumn{1}{l|}{346.1} & 346.72 & caring & 0.454 & 0     & No face detected \\
\cmidrule{2-10}          & 132   & \multicolumn{2}{p{20.915em}|}{" I mean, I guess I like to see some plus video of like here either going one path or"} & \multicolumn{1}{l|}{557.34} & 562.74 & neutral & 0.251 & 0     & No face detected \\
\cmidrule{2-10}          & 133   & \multicolumn{2}{p{20.915em}|}{another.} & \multicolumn{1}{l|}{562.74} & 563.44 & neutral & 0.959 & 0     & No face detected \\
\cmidrule{2-10}    \multicolumn{1}{c}{} & \multicolumn{1}{l}{} & \multicolumn{1}{l}{} & \multicolumn{1}{l}{} &       & \multicolumn{1}{l}{} & \multicolumn{1}{l}{} & \multicolumn{1}{l}{} & \multicolumn{1}{l}{} & \multicolumn{1}{r}{} \\
    \midrule
    \rowcolor[rgb]{ .949,  .949,  .949} \multicolumn{1}{|c|}{\textbf{5}} & \multicolumn{9}{l|}{\textbf{Can you find anomalies in speech and facial emotions?}} \\
    \midrule
          & \textbf{index} & \multicolumn{4}{p{35.415em}|}{\textbf{text}} & \textbf{speech\_emotion} & \textbf{facial\_emotion} & \textbf{avg\_fer\_score} & \textbf{confidence} \\
\cmidrule{2-10}          & 2     & \multicolumn{4}{p{35.415em}|}{And that Own Attent.} & neutral & angry & 0.736 & 0.661 \\
\cmidrule{2-10}          & 6     & \multicolumn{4}{p{35.415em}|}{" Yeah, this is the Tent."} & neutral & happy & 0.994 & 0.461 \\
\cmidrule{2-10}          & 7     & \multicolumn{4}{p{35.415em}|}{Sorry.} & remorse & happy & 0.995 & 0.399 \\
\cmidrule{2-10}          & 8     & \multicolumn{4}{p{35.415em}|}{Now I'm going to write a description about my Tent.} & neutral & happy & 0.955 & 0.693 \\
\cmidrule{2-10}          & 11    & \multicolumn{4}{p{35.415em}|}{" Two, and if you move a lot in your sleep, it's pretty light."} & caring & neutral & 0.863 & 0.284 \\
\cmidrule{2-10}          & 13    & \multicolumn{4}{p{35.415em}|}{Cool.} & admiration & neutral & 0.577 & 0.361 \\
\cmidrule{2-10}          & 14    & \multicolumn{4}{p{35.415em}|}{And I'm going to set my price.} & neutral & angry & 0.58  & 0.45 \\
\cmidrule{2-10}          & 17    & \multicolumn{4}{p{35.415em}|}{It's like \$100 Tent.} & neutral & sad   & 0.676 & 0.824 \\
\cmidrule{2-10}          & 18    & \multicolumn{4}{p{35.415em}|}{That would be pretty cool.} & admiration & neutral & 0.732 & 0.665 \\
\cmidrule{2-10}          & 19    & \multicolumn{4}{p{35.415em}|}{I'll pick up Address.} & neutral & sad   & 0.714 & 0.692 \\
\cmidrule{2-10}          & 20    & \multicolumn{4}{p{35.415em}|}{Where will people pick it up from you?} & curiosity & angry & 0.64  & 0.474 \\
\cmidrule{2-10}          & 24    & \multicolumn{4}{p{35.415em}|}{What's this?} & curiosity & neutral & 0.96  & 0.488 \\
\cmidrule{2-10}          & 26    & \multicolumn{4}{p{35.415em}|}{I'm going to click here.} & neutral & angry & 0.53  & 0.496 \\
\cmidrule{2-10}          & 28    & \multicolumn{4}{p{35.415em}|}{I put three.} & neutral & angry & 0.603 & 0.93 \\
\cmidrule{2-10}          & 30    & \multicolumn{4}{p{35.415em}|}{But I'll put three again.} & neutral & sad   & 0.772 & 0.7 \\
\cmidrule{2-10}          & 31    & \multicolumn{4}{p{35.415em}|}{And I'm going to post my listing.} & sadness & neutral & 0.717 & 0.18 \\
\cmidrule{2-10}          & 32    & \multicolumn{4}{p{35.415em}|}{Sweet.} & admiration & neutral & 0.575 & 0.298 \\
\cmidrule{2-10}          & 34    & \multicolumn{4}{p{35.415em}|}{So now I'm just going to double check that my post looks good.} & approval & neutral & 0.646 & 0.248 \\
\cmidrule{2-10}          & 35    & \multicolumn{4}{p{35.415em}|}{Awesome.} & admiration & sad   & 0.843 & 0.922 \\
\cmidrule{2-10}          & 36    & \multicolumn{4}{p{35.415em}|}{Cool.} & admiration & neutral & 0.665 & 0.361 \\
\cmidrule{2-10}          & 37    & \multicolumn{4}{p{35.415em}|}{Thanks.} & gratitude & neutral & 0.798 & 0.993 \\
\cmidrule{2-10}          & 38    & \multicolumn{4}{p{35.415em}|}{Cool.} & admiration & neutral & 0.844 & 0.361 \\
\cmidrule{2-10}          & 40    & \multicolumn{4}{p{35.415em}|}{And I'm just going to kind of get us back to the homepage so we can restart.} & neutral & No face detected & 0     & 0.255 \\
\cmidrule{2-10}    \multicolumn{1}{c}{} & \multicolumn{1}{l}{} & \multicolumn{1}{l}{} & \multicolumn{1}{l}{} &       & \multicolumn{1}{l}{} & \multicolumn{1}{l}{} & \multicolumn{1}{l}{} & \multicolumn{1}{l}{} & \multicolumn{1}{r}{} \\
    \midrule
    \rowcolor[rgb]{ .949,  .949,  .949} \multicolumn{1}{|c|}{\textbf{6}} & \multicolumn{9}{l|}{\textbf{If I were to map the speech emotions into facial emotions, were there any anomalies where speech emotions did not match facial emotions?}} \\
    \midrule
          & \textbf{index} & \multicolumn{2}{p{20.915em}|}{\textbf{text}} & \multicolumn{2}{l|}{\textbf{emotion}} & \multicolumn{2}{l|}{\textbf{mapped\_fer\_emotion}} & \multicolumn{2}{l|}{\textbf{dominant\_fer\_emotion}} \\
\cmidrule{2-10}          & 0     & \multicolumn{2}{p{20.915em}|}{Sorry.} & \multicolumn{2}{l|}{remorse} & \multicolumn{2}{l|}{sad} & \multicolumn{2}{l|}{happy} \\
\cmidrule{2-10}          & 1     & \multicolumn{2}{p{20.915em}|}{Where will people pick it up from you?} & \multicolumn{2}{l|}{curiosity} & \multicolumn{2}{l|}{surprise} & \multicolumn{2}{l|}{angry} \\
\cmidrule{2-10}          & 2     & \multicolumn{2}{p{20.915em}|}{What’s this?} & \multicolumn{2}{l|}{curiosity} & \multicolumn{2}{l|}{surprise} & \multicolumn{2}{l|}{neutral} \\
\cmidrule{2-10}          & 3     & \multicolumn{2}{p{20.915em}|}{So now I’m just going to double check that my post looks good} & \multicolumn{2}{l|}{approval} & \multicolumn{2}{l|}{happy} & \multicolumn{2}{l|}{neutral} \\
\cmidrule{2-10}          & 4     & \multicolumn{2}{p{20.915em}|}{Is there any foot salad?} & \multicolumn{2}{l|}{curiosity} & \multicolumn{2}{l|}{surprise} & \multicolumn{2}{l|}{happy} \\
\cmidrule{2-10}          & 5     & \multicolumn{2}{p{20.915em}|}{Alright.} & \multicolumn{2}{l|}{approval} & \multicolumn{2}{l|}{happy} & \multicolumn{2}{l|}{neutral} \\
\cmidrule{2-10}          & 6     & \multicolumn{2}{p{20.915em}|}{Sorry.} & \multicolumn{2}{l|}{remorse} & \multicolumn{2}{l|}{sad} & \multicolumn{2}{l|}{happy} \\
\cmidrule{2-10}          & 7     & \multicolumn{2}{p{20.915em}|}{Was there anything that was hard or frustrating?} & \multicolumn{2}{l|}{curiosity} & \multicolumn{2}{l|}{surprise} & \multicolumn{2}{l|}{sad} \\
\cmidrule{2-10}          & 8     & \multicolumn{2}{p{20.915em}|}{Because it kind of looks like a different thing, I guess.} & \multicolumn{2}{l|}{approval} & \multicolumn{2}{l|}{happy} & \multicolumn{2}{l|}{sad} \\
\cmidrule{2-10}          & 9     & \multicolumn{2}{p{20.915em}|}{The price, yeah, one the one-sword, it’s got a dark picture.} & \multicolumn{2}{l|}{approval} & \multicolumn{2}{l|}{happy} & \multicolumn{2}{l|}{sad} \\
\cmidrule{2-10}          & 10    & \multicolumn{2}{p{20.915em}|}{Yeah, that’s a stretch goal.} & \multicolumn{2}{l|}{approval} & \multicolumn{2}{l|}{happy} & \multicolumn{2}{l|}{sad} \\
\cmidrule{2-10}          & 11    & \multicolumn{2}{p{20.915em}|}{Okay, so the feature, I guess, I really valued.} & \multicolumn{2}{l|}{approval} & \multicolumn{2}{l|}{happy} & \multicolumn{2}{l|}{sad} \\
\cmidrule{2-10}    
\end{tabular}%
}

\label{tab:addlabel}
\end{table}

\begin{table}[htbp]
  \centering
  \resizebox{\textwidth}{!}{
    \begin{tabular}{c|l|p{13.75em}|l|l|l|l|l|l|l|}
    \toprule
    \rowcolor[rgb]{ .949,  .949,  .949} \multicolumn{1}{|c|}{\textbf{7}} & \multicolumn{9}{l|}{\textbf{Can you find where the user recommended any improvement or would like to have something different from what was presented?}} \\
    \midrule
          & \textbf{index} & \textbf{text} & \textbf{start} & \textbf{end} & \textbf{emotion} & \textbf{confidence} & \textbf{avg\_fer\_score} & \textbf{dominant\_fer\_emotion} & \textbf{anomaly} \\
\cmidrule{2-10}          & 16    & Probably just put it up for like \$5 per day. & 97.06 & 101.04 & neutral & 0.663 & 0.725 & neutral & FALSE \\
\cmidrule{2-10}          & 17    & It's like \$100 Tent. & 101.42 & 102.7 & neutral & 0.824 & 0.676 & sad   & TRUE \\
\cmidrule{2-10}          & 23    & Like 10. & 117.78 & 118.28 & neutral & 0.887 & 0.955 & neutral & FALSE \\
\cmidrule{2-10}          & 56    & It looks like this is a two person tent. & 239.9 & 244.72 & neutral & 0.598 & 0.834 & neutral & FALSE \\
\cmidrule{2-10}          & 68    & It has like results overlapped with this. & 307.42 & 312.08 & neutral & 0.758 & 0.76  & neutral & FALSE \\
\cmidrule{2-10}          & 69    & But it went away like halfway and clicked something else. & 312.58 & 315.66 & neutral & 0.573 & 0.742 & neutral & FALSE \\
\cmidrule{2-10}          & 98    & " because it kind of looks like a different thing, I guess." & 405.88 & 409.46 & approval & 0.282 & 0.761 & sad   & TRUE \\
\cmidrule{2-10}          & 101   & " Yeah, but you can maybe make that like an icon in itself or something." & 427.9 & 431.66 & neutral & 0.405 & 0.86  & sad   & TRUE \\
\cmidrule{2-10}          & 106   & " Look at the next tent, because I look at this and I'd be like, " & 455.88 & 459.4 & neutral & 0.298 & 0.784 & sad   & TRUE \\
\cmidrule{2-10}          & 111   & " Other than that, yeah, it looks like this is a person." & 473.28 & 475.88 & neutral & 0.416 & 0.801 & sad   & TRUE \\
\cmidrule{2-10}          & 115   & It's like a rating system. & 483.84 & 485.1 & neutral & 0.792 & 0.647 & sad   & TRUE \\
\cmidrule{2-10}          & 122   & would be like a Lister Uber. & 507.4 & 509.26 & neutral & 0.63  & 0.678 & neutral & FALSE \\
\cmidrule{2-10}          & 127   & " You know, like Drew Persons." & 538   & 539.2 & neutral & 0.438 & 0.903 & neutral & FALSE \\
\cmidrule{2-10}    \multicolumn{1}{c}{} & \multicolumn{1}{l}{} & \multicolumn{1}{l}{} & \multicolumn{1}{l}{} & \multicolumn{1}{l}{} & \multicolumn{1}{l}{} & \multicolumn{1}{l}{} & \multicolumn{1}{l}{} & \multicolumn{1}{l}{} & \multicolumn{1}{r}{} \\
    \midrule
    \rowcolor[rgb]{ .949,  .949,  .949} \multicolumn{1}{|c|}{\textbf{8}} & \multicolumn{9}{l|}{\textbf{This data is based on user testing, as part of user experience study, can you tell me the key places where the experience can be improved?}} \\
    \midrule
          & \textbf{index} & \multicolumn{2}{p{20.915em}|}{\textbf{text}} & \textbf{start} & \textbf{end} & \textbf{emotion} & \textbf{confidence} & \textbf{avg\_fer\_score} & \textbf{dominant\_fer\_emotion} \\
\cmidrule{2-10}          & 50    & \multicolumn{2}{p{20.915em}|}{I'm not really sure if I should click the menu looking thing or take a icon.} & 207   & 214.26 & confusion & 0.444 & 0.582 & neutral \\
\cmidrule{2-10}          & 63    & \multicolumn{2}{p{20.915em}|}{I'm not really sure how to find.} & 277.82 & 279.86 & confusion & 0.655 & 0.497 & neutral \\
\cmidrule{2-10}          & 92    & \multicolumn{2}{p{20.915em}|}{I clicked it and then I wasn't really sure exactly how to filter three results.} & 376.54 & 382.46 & confusion & 0.381 & 0.785 & sad \\
\cmidrule{2-10}          & 93    & \multicolumn{2}{p{20.915em}|}{I'm not sure what that was.} & 384.84 & 387.62 & confusion & 0.648 & 0.633 & sad \\
\cmidrule{2-10}          & 95    & \multicolumn{2}{p{20.915em}|}{I'm not really sure these three I can't do.} & 391.46 & 393.54 & confusion & 0.587 & 0.725 & neutral \\
\cmidrule{2-10}          & 121   & \multicolumn{2}{p{20.915em}|}{" So I'm never used Airbnb or anything, but I guess the most similar product I have used"} & 499.8 & 507.4 & confusion & 0.296 & 0.727 & neutral \\
\cmidrule{2-10}    \end{tabular}%
}
\captionsetup{justification = centering}
\caption{The Q/A session through PandasAI conversational tool.}
  \label{tab:tb6}%
\end{table}%

 Knowing when the user felt happy helps to capture the strengths of the product. In Testing\_video1, there were instances detected where the user was happy. Due to data integration, it is now possible to extract insights about the timing and part of the video when this happened. In Q2 in Table \ref{tab:tb6}, it can be seen that the user was happy at 330.08 seconds in the video with the statement and the confidence score of the text. \par

Confusion or frustration are emotions that depicts when a person is not able to understand the task or felt annoying. Insights into these emotions help to find areas of improvement for UX designers. The response in Table \ref{tab:tb6} Q3 illustrates instances in which the user was unable to perform tasks conveniently. \par

Statements made by the user such as “I’m not really sure how to find.”, “I clicked it and then I wasn’t really sure exactly how to filter three results.” are meaningful insights for a UX designer that would allow them to track down the time when the task was being performed and which task was being performed that confused the user. From the above figure, it can also be observed that the facial emotion is sad and the speech emotion is confusion, align with each other, making a strong impression of the statement and providing an example of benefits of the data integration step performed earlier. \par

Sometimes, it is possible that data might be lost or not recorded, the simple question asked about whether there were instances when no facial expressions were recorded was made to investigate the same problem. In Table \ref{tab:tb6} Q4, it can be seen that there are some instances where the face was not detected in the video, but since there is speech emotion recorded, it might not affect the insight gathering process to a greater extent, thus again demonstrating integration of data from different modalities as a wise method for insight gathering. \par

One of the objectives of the project was to find anomalies through the implementation of different modalities for deeper insight extractions. LLM was able to find instances where the facial and speech emotions did not align as shown in Table \ref{tab:tb6} Q5 response. A total of 100 anomalies were found. \par

To further refine the response, a question was asked to which the LLM provided more detailed answer by displaying a generated column called ‘mapped\_fer\_emotion’ as can be seen in Table \ref{tab:tb6} Q6. The LLM was able to manipulate the data itself and generate the desired response. \par



Detecting user speech has a very important benefit for the UX designer. Any suggestions, feedback and advice help to improve the service. Designers had to ask for feedback or conduct surveys after user testing. Keeping this in mind, the next question about instances in which user recommendations were requested and the response can be seen in Table \ref{tab:tb6} Q7.\par


However, the suggestion made by the user as seen in index 101 is displayed, but with other irrelevant information, shows limitations in handling complex questions. \par

Through the final question in Table \ref{tab:tb6} Q8, it was observed that the current LLM is capable of generating responses to questions in which any data in the file are not explicitly mentioned. \par


According to the findings, the designer should focus on areas or duration when the user was confused or annoyed. A timeline visualization is also presented, illustrating the key highlights and insights found through the implementation of automating user emotion detection from Testing\_video1. \par
\begin{figure}[H]
    \centering
    \includegraphics[width=\linewidth]{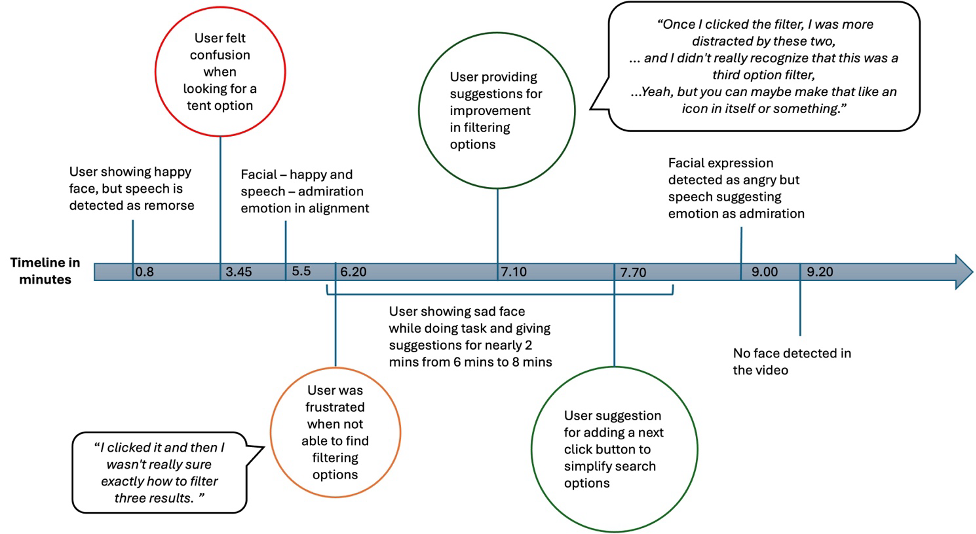}
    \captionsetup{justification=centering}
    \caption{Timeline visualization of Testing\_video1.}
    \label{fig:fg27}
\end{figure}

\section{Conclusion}
Recalling the purpose and goal of the project, the project aimed in getting emotional insights from a user research video. To achieve this aim, the project was done following SMART objectives. Firstly, the importance of emotions in a person’s life and recognizing other people’s emotions were discussed. Different emotion theories and models were explored to get a grasp of what emotions are and what are the mechanics behind it. The need to recognize emotions could be found in various applications such as healthcare, medical, customer services and HCI. \par
With the help available technology, it is now possible to automate traditional ways of working which were time-consuming and tedious. In the field of HCI, user testing or UX evaluation plays an important role in gathering insights for fulfilling the user’s needs and expectations. A UX designer needs feedback and suggestions from its customers to improve the service and meet expectations. One way of doing this is through user testing in which a task is assigned to a participant/user, and the user talks about the strengths and weaknesses he/she experienced while reviewing the product in a recorded session. The role of UX designer is to gather insights from the user testing video and keep track of user’s changing emotions, key areas where the experience need to be improved, user’s recommendations made while reviewing, etc. Traditionally, to meet all these requirements, designers had to re-watch the whole video again and again to keep track and note everything. And after performing this time-consuming and tedious task, they had to perform data analysis for their key findings. With the help of AI, and the solution presented in this project , it was made possible to automate the evaluation process of extracting insights from a video. \par
Insights were gathered from different emotion recognition modalities such as facial emotion recognition, speech-to-text and text-based emotion recognition. For FER process, a pre-trained ViT based model trained on FER-2013 dataset was implemented which was chosen on the basis of quantitative as well as qualitative analysis. Among 10 available pre-trained ViT models, the selected model performed well on three facial recognition benchmark datasets i.e. FER-2013, CK+48 and AffectNet, displaying better generalization ability compared to other models. For speech-to-text conversion, Open AI's Whisper model was chosen and for text-based emotion recognition process a pre-trained BERT based model trained on GoEmotions dataset from Google with evaluation accuracy of 96.14\% was implemented. \par
The FER model recognized emotions into 7 basic emotions of sad, happy, neutral, angry, surprise, disgust and fear whereas the text-based ED model recognized 27 different emotions. For better understanding, the data integration step was implemented. Not only this was necessary for gathering richer data for analysis, but it was also equally important for UX evaluation. 
Data integration made it possible to extract deeper meaningful insights by combining data from different modalities. This step also helped finding any anomalies or patterns depicted in user behaviour. \par
Data integration was done using Temporal alignment and Temporal Fusion methods in which the data from different modalities was aligned based on timestamps and this temporally aligned data was then fused based on time. Finally, the integrated data was fed to a conversational data analyst tool i.e. PandasAI, which allows incorporating different LLMs, through which the designer could interact with the data in a natural language way without the need of learning SQL queries. This would reduce the burden of manually processing the data and make the process of UX evaluation time-efficient and convenient. Among available options, Open AI's GPT- 4o was used for data analysis process in which the designer could directly ask questions about the user testing video. In this process, the tool provided answers focused on the video data provided. User testing is a confidential process among organizations due to sensitive nature of emotional data. Directly uploading a video on any general purpose LLM would invade privacy and integrity of the organization/company, therefore, a more secure way of data processing would be beneficial.\par

In addition, general-purpose LLMs generate generic answers. In this project, the use of PandasAI enabled the incorporated LLM to focus only on the data provided and uncover specific trends and actionable insights that a general-purpose LLM might not provide. The integration of data enables correlation with facial expression and spoken content, which might be difficult to achieve with a general purpose LLM. \par
By integrating facial and speech data, the project contributes to the research of multimodal analysis, offering a comprehensive approach to understanding user emotions. Incorporating a conversational data analyst tool, the project showcases a novel application of AI in UX research and evaluation, contributing to methodological advancements in the field. \par
The gained insights on user emotions directly inform and improve UX design. Automating emotion detection and data processing reduces the need for manual review, making user testing more scalable and efficient for UX designers. In addition, to the researcher’s knowledge, very limited applications were found where a user can extract any sort of insight from a video. This project is a step forward in its application in the field of UX. \par
\subsection{Future developments and scope}
More accurate and efficient emotion recognition models are required for better application. Among all the challenges and limitations, a successful application of emotion recognition and data processing technology was implemented in the field of UX. Future work would include better models where different speakers could be recognized and insights could be gathered accordingly. Real-time emotion detection and analysis could be implemented to get immediate feedback on UX improvements. In addition to face and speech emotion recognition, physiological sensors could provide more deeper insight. The proposed workflow could be implemented in other fields of applications such as education, healthcare, and marketing to extract insights from a video. An application could be implemented for a better demonstration of the project. 
\subsection{Ethical Considerations}
To prevent unfair and inaccurate detection of user emotions in FER model, the models were evaluated on three different dataset representing different age groups and gender. The model with the most generalizability was chosen for implementation. Detecting user emotions from facial and speech expressions is sensitive information; therefore, only publicly available videos were used to demonstrate the concept to prevent any breach of privacy and consent.




\bibliographystyle{abbrv}
\bibliography{library}


\end{document}